\documentclass[prd, onecolumn,showpacs,nofootinbib,preprintnumbers,superscriptaddress]{revtex4-2}

\usepackage{amsmath}\usepackage{amsfonts}\usepackage{graphicx}\usepackage{hyperref}\usepackage{mathbbol}\usepackage{multirow} 
\def\be{\begin{equation}}   
\def\ee{\end{equation}}  
 \def\bea{\begin{eqnarray}}    \def\eea{\end{eqnarray}}     \def\f{\frac} 
    \def\d{{\rm d}}   \def\m{{\rm m}}      \def\k{\kappa}       
\def\r{\right}            \def\l{\left}
\def\ve{\varepsilon}

\begin{document}
\title{Dynamics of holographic dark energy with apparent-horizon cutoff and non-minimal derivative coupling gravity in non-flat FLRW universe}
\date{\today}

 \author{Amornthep Tita}\email{amornthepti61@nu.ac.th}
 \affiliation{The Institute for Fundamental Study ``The Tah Poe Academia Institute", Naresuan University, Phitsanulok 65000, Thailand}
  \affiliation{NAS, Centre for Theoretical Physics \& Natural Philosophy, Mahidol University, Nakhonsawan Campus,  Phayuha Khiri, Nakhonsawan 60130, Thailand}
\author{Burin Gumjudpai}\email{Corresponding: burin.gum@mahidol.ac.th}
\affiliation{NAS, Centre for Theoretical Physics \& Natural Philosophy, Mahidol University, Nakhonsawan Campus,  Phayuha Khiri, Nakhonsawan 60130, Thailand}
 
  \author{Pornrad Srisawad}\email{pornrads@nu.ac.th}
  \affiliation{Department of Physics, Faculty of Science, Naresuan University, Phitsanulok 65000, Thailand}

\begin{abstract}
Background cosmological dynamics for a universe with matter, a scalar field non-minimally derivative coupling to Einstein tensor under power-law potential and holographic vacuum energy is considered here. The holographic IR cutoff scale is apparent horizon which, for accelerating universe, forms a trapped null surface in the same spirit as blackhole's event horizon. For non-flat case, effective gravitational constant cannot be expressed in the Friedmann equation. Therefore holographic vacuum density is defined with standard gravitational constant instead of the effective one.  Dynamical and stability analysis shows four independent fixed points. One fixed point is stable and it corresponds to $w_{\text{eff}} = -1$. 
One branch of the stable fixed-point solutions corresponds to  de-Sitter expansion. 
The others are either unstable or saddle nodes. 
Numerical integrations of the dynamical system are performed and plotted confronting with $H(z)$ data. 
It is found that for flat universe, $H(z)$ observational data favors large negative value of NMDC coupling, $\kappa$. Larger holographic contribution, $c$, and larger negative NMDC coupling increase slope and magnitude of the $w_{\text{eff}}$ and $H(z)$. Negative $\kappa$,  can contribute to phantom equation of state, $w_{\text{eff}} < -1$. The NMDC-spatial curvature coupling could have phantom energy contribution. Free negative spatial curvature term can also contribute to phantom equation of state, but only with significantly large negative value of the spatial curvature.   
The model could give phantom equation of state for $\kappa = -200$  and high value of $c$ for both flat and open cases.  
             
\end{abstract}

\date{\today}
\vskip 1pc
\maketitle \vskip 1pc


\section{Introduction}  \label{sec_intro}
Present acceleration is a puzzle of contemporary cosmology. Dark energy or modification of general relativity could result in the acceleration \cite{CopelandDynamics, Paddy2006, DEBook, VF,CL,CF, odin,en, Ishak:2018his}. The acceleration corresponds to negative equation of state, $w < -1/3$ and  the observational favored value  is  $w \approx -1$   \cite{Amanullah2010, Astier:2005qq, Goldhaber:2001a, Perlmutter:1997zf, Perlmutter:1999a, Riess:1998cb, Scranton:2003, Tegmark:2004}. 
Dark energy (DE) is hypothetical energy with repulsive pressure. It could be cosmological constant or dynamical  scalar fields. Having dark energy content in the universe is  equivalent to  adding of extra degree of freedom to the matter Lagrangian.  Alternative way of achieving late acceleration is the modification of general relativity, i.e. modifying the left side of the Einstein field equation, that is, the gravitational sector.  
There are many ways of gravitational modifying such as considering function of Ricci scalar 
\cite{DeFelice:2010aj}, function of Ricci tensor and Riemann tensor instead of using the Einstein-Hilbert Lagrangian \cite{Carroll2004}.  Many other models are of mixed types that allow couplings among barotropic fluid, scalar and gravitational sectors. As a result, there are  rich implications of these couplings in scalar-tensor theories  \cite{BD1961, Maeda, VF}   
such as mediation of  long-range fifth force when coupling between matter and scalar field is allowed. In this case, chameleon screening mechanism is considered to evade the fifth force problem  \cite{charm}.  
It is found that only
 $R \phi_{,\mu}\phi^{,\mu}$
and  $R^{\mu\nu} \phi_{,\mu} \phi_{,\nu}$  terms are necessary in the coupling sector \cite{Amendola1993, Capozziello:1999xt}. These couplings are motivated in lower energy limits of higher dimensional
theories or in conformal supergravity
\cite{Liu:1998bu, Nojiri:1998dh}. Combining the two terms into the Einstein tensor coupling to derivative of scalar field as $G_{\mu \nu} (\nabla^{\mu}\phi )(\nabla^{\nu}\phi)$ gives rise to the non-minimal derivative coupling (NMDC) gravity model and it can result in de-Sitter expansion as seen in iteratures     
\cite{Capozziello:1999uwa, Granda:2010hb, Granda:2010ex, Granda:2011zk, Sushkov:2009, Saridakis:2010mf, Gao:2010vr, Germani:2010gm, Sushkov:2012, Skugoreva:2013ooa,  Matsumoto:2015hua, Koutsoumbas:2013boa, Darabi:2013caa, Germani:2009, Germani:2010hd, Dalianis:2016wpu, Sadjadi:2012zp, Tsujikawa:2012mk, Ema:2015oaa, Jinno:2013fka, Ema:2016hlw, Yang:2015pga, Sadjadi:2010bz, Gumjudpai:2015vio}. 
 Further generalization of scalar-tensor theories, with at most second-order derivative with respect to dynamical variables, are  such as galileons \cite{Nicolis:2008in, Deffayet:2009wt, Deffayet:2009mn}, Fab-Four \cite{Charmousis:2011bf}, Horndeski action \cite{Horn, Deffayet:2011, Kobayashi} and GLPV theories \cite{GLPV}. 
   The $G_{\mu \nu}(\nabla^{\mu}\phi)(\nabla^{\nu}\phi)$  term is a sub-class (the $G_5$ term) of the Horndeski action.

 Action of the NMDC coupling to gravity with other matters, e.g. dark matter (DM), matter fields and cosmological constant, is given by
   \be
   S=\int \d^4x\sqrt{-g}\left\{\frac{R}{16\pi G}-\frac{[\varepsilon g_{\mu\nu}+\kappa G_{\mu\nu}]}{2} (\nabla^{\mu}\phi) (\nabla^{\nu}\phi) - V(\phi)   \right\}  + S_{\rm m, \Lambda}.    \label{NMDCaction}
   \ee
   The parameter $\varepsilon$ is $\pm 1$ for canonical and phantom cases. The coupling constant $\k$ has $mass^{-2}$ dimension. 
  Sign of the coupling $\k$ could either enhance or reduce contribution of  the free kinetic term  \cite{Granda:2010hb} which in turn affects  power spectral index, tensor-to-scalar ratio, evolution of the equation of state and other observational parameters  \cite{Caldwell:2003vq, Saridakis:2010mf, Skugoreva:2013ooa, Quiros:2017gsu, Baisri:2022ivv}.
Observational data has put tight constraints on form of scalar potentials for viability of the NMDC inflationary model in metric formalism \cite{Tsujikawa:2012mk, Yang:2015pga} 
while, in Palatini formalism, the quatic power-law inflationary potential is completely ruled out by the CMB data \cite{Gumjudpai:2016ioy, Saichaemchan:2017psl, Muhammad:2018dwi}.
NMDC inflation, if the initial scalar field speed is sufficiently fast, can end up with quasi-de Sitter expansion with graceful exits. Even without scalar potential, $V(\phi)$, which is to be converted to kinetic energy, quasi-de Sitter expansion with graceful exits can still happen. 
 For a power-law inflationary potential  $V(\phi) = V_0 \phi^n$ with $n=1.5$, double inflation, i.e. kinetic driving and potential driving, with sub-Planckian initial scalar field value can solve the horizon problem with large coupling and very small scalar mass \cite{Avdeev:2021von}. 
All of these are good aspects of the NMDC inflationary model.  At last, shortcomings of the NMDC inflation are the prediction of too large tensor-to-scalar ratio and the absence  of graceful exits for Higgs-like potential or for $n \leq 2$   \cite{Matsumoto:2017gnx, Granda:2017dlx}. NMDC inflation with power-law potential is hence disfavored by CMB data in this setting.  At late time, the model could result in $w \rightarrow -1$  
 for a power-law potential  $V(\phi) = V_0 \phi^n$  with  $n \leq 2$ ending with oscillating scale factor and it results in Big Rip singularity for $n > 2$ \cite{Sushkov:2009, Bruneton:2012zk, Sushkov:2012, Skugoreva:2013ooa}. Other potentials such as Higgs-like and exponential potentials for the case have been studied \cite{Matsumoto:2015hua}. A detail qualitative study of NMDC cosmological dynamics is reported \cite{Matsumoto:2017gnx}.

In searching for unifying theory of gravitation and quantum behavior, a compelling holographic principle comes naturally to exist \cite{tHooft:1993dmi}. Following this, Susskind describes the principle in context of string theory \cite{Susskind}. Maldacena proposes AdS/CFT correspondence which views conformal field theory on the surface of a bulk region as hologram of corresponding string theory in the bulk  \cite{Maldacena:1997re}. Surface area enclosing volumic bulk is linked to entropy of the bulk and this is known as Bekenstein-Hawking entropy \cite{Be1, Be2, Be3, Hawking, Hawking2}. Since any surface area can be mostly sub-divided to the smallest area in Planck scale hence there is limited information of quantum states of the surface area. Therefore, there must be an maximum entropy bound for a bulk region  \cite{Bousso:1999dw, Bousso:2002bh}. A blackhole is created when information exceeds the entropy bound. For a blackhole,  its entropy is proportional to surface area $S \sim A/4G$  or  square of length scale of a blackhole $L_{\rm BH}^2$. 
 Blackhole hence contains holographic information on its event horizon \cite{Bousso:2002ju}.
  Applying the entropy bound hypothesis to cosmology, equation of state should be bounded within $w< 1$ and the universe should be infinite  \cite{Fischler:1998st}
  In this view, there is a relation  $\rho_{\Lambda} \propto S L^{-4}$ between UV energy scale ($\rho_{\Lambda}$), and IR cosmic length scale ($L$) 
 \cite{Cohen, Thomas, Horava, Hsu}.  At surface of boundary, there is a hologram of information in the cosmic bulk. This implies
 \begin{eqnarray}  \label{1}
\rho_{\Lambda} &=&  \frac{3c^2}{8\pi GL^2},
\end{eqnarray}
as IR cutoff to cosmological constant density. The factor $c$ is a constant $ (0 \lesssim c <1)$ \cite{Li}. Hence this could solve fine-tuning problem. One support for the holographic principle is that the Casimir energy is found to be proportional to the horizon size \cite{Li:2009pm}. 
When considering domination of cosmological constant in the universe and the length scale is Hubble horizon, $L \sim H^{-1}$, i.e. flat case of apparent horizon, dark energy equation of state is dust-like, $w \approx 0 $ which is not acceptable \cite{Hsu}. 
Alternatively, using particle horizon as cutoff IR length scale (\cite{Fischler:1998st, Bousso:1999xy}) gives $w > -1/3$ \cite{Li}. 
To obtain accelerating universe with acceptable equation of state, $w < -1/3 $, future event horizon IR cutoff is considered   \cite{Li, Pavon}.  
Using  future event horizon can also solve cosmic coincidence 
problem for at least $N> 60$ inflationary e-folding number.  With future event horizon cutoff, since phantom equation of state is observationally allowed and     
it violates the second law of thermodynamics  \cite{Gong:2006sn}. 
Existing of turning point of Hubble parameter also violates the Null Energy Condition because, with $c < 1$, the future event horizon cutoff model results in $w < -1$ \cite{Colgain:2021beg}. 
In order to alleviate these problems, interaction between DM and DE is introduced so that effective equation of state can cross the phantom barrier. The interaction can also solve of the cosmic coincidence problem  \cite{Pavon, Karwan:2008ig}. However the original holographic dark energy (HDE) model with future event horizon cutoff in flat universe suffers from cosmic age problem, that is, it predicts a universe with younger age than those of high-$z$ objects unless forcing $h \lesssim 0.56$ \cite{Wei:2007ig} although this can be slightly avoidable in non-flat case  \cite{Cui:2010dr} or allowing DM-DE interaction. Later a few newer cutoff scales have been proposed such as  agegraphic holographic dark energy \cite{Cai:2007us, Wei:2007ty, Wei:2007xu} and Ricci holographic dark energy which have been ruled out by observations  \cite{Zhang:2009un, LiLiWang2009, XuZhang2016}. The other models of cutoff length scale are such as Granda and Oliveros
\cite{Granda:2008dk, Granda:2008tm} and other models as seen in  \cite{Chen:2009zzv, Chattopadhyay:2020mqj,  Chattopadhyay:2020xx, Chattopadhyay:2020x2, Nojiri:2021iko}. The Granda and Oliveros cutoff model cannot satisfy the expansion data when combined with perturbation data  \cite{Akhlaghi:2018knk}. 
Good features and many problems can be cured with time-varying $c$ \cite{Malekjani:2018qcz} but this is to add a new parameter to the theory.  Review on HDE models can be seen in \cite{Wang:2016och}.

Cosmological horizon of the IR cutoff  scale should have similar characters to blackhole's event horizon.
Cosmic bulk volume should be enclosed by a trapped null surface, of which horizon can never be reached by light signal. 
In a universe under cosmic acceleration, trapped null surface exists when using apparent horizon cutoff scale such that light signal can never reach the apparent horizon. 
Connection of the first law of thermodynamics to the Friedmann equation \cite{Cai:2005ra} suggests definition of Cai-Kim temperature, 
$ T = 1/(2 \pi R_{\rm A}) $  
defined with the size of apparent horizon,
$ R_{\rm A} =  1/\sqrt{H^2 + k/a^2}$.  When the curvature, $k$, is $0$, the apparent horizon is the Hubble length. While the  
 HDE is usually considered to be vacuum energy, scalar field could also be negative pressure cosmic component.  It has been known that, to solve the present phantom-crossing problem, one cannot use single canonical scalar field model nor some k-essence models (due to no-go theorem for the k-essence case) \cite{Vikman:2004dc}.
Quintom models composed of both quintessence and phantom kinetic terms \cite{Feng:2004ad} are able to result in phantom crossing, however they usually process one ghost degree of freedom \cite{Cai:2009zp}. Avoiding the ghost, complex scalar field versions of quintom are considered however suffering from Q-ball formation.  Hessence model, a quintom-like model with non-canonical complex scalar field, can cure the Q-ball problem with conservative charge of the theory \cite{Wei:2005nw}. The hessence model is considered in holographic scenario where the hessence scalar field DE density is cutoff by future event horizon. This model can have phantom-crossing behavior and it is known as holographic hessence model \cite{Zhao:2007qy}.

In context of scalar-tensor theories, a unification of inflation and late-time phantom-crossing can be possible when considering dilaton-like self-coupling of scalar kinetic term or with generalized version of the holographic cutoffs \cite{Nojiri:2005pu}.  
In Brans-Dicke gravity (Jordan frame), using Hubble scale cutoff and particle horizon cutoff
in original HDE model cannot accelerate the expansion while using future event horizon cutoff can achieve the acceleration \cite{Gong:2004fq}.  
Another HDE model with Hubble scale cutoff in scalar-tensor theories, allowing DM-DE interaction, can achieve deceleration-to-acceleration transition \cite{Bisabr:2008gu}.
In Brans-Dicke gravity, the scalar potential $V(\phi)$ is necessary for the HDE model with Hubble scale cutoff to be viable  \cite{Liu:2009ha, Gong:2008br}. There are also other interests of scalar field non-minimal coupling to gravity (NMC) term in HDE cosmology with Hubble scale cutoff \cite{Ito2005, Setare:2008pc, Granda:2009zx}. Considering NMDC model, HDE NMDC flat cosmology with Hubble scale cutoff has been investigated \cite{Kritpetch:2020vea}. However, for power-law and exponential potential, the model gives mismatched inflationary parameters \cite{Baisri:2022ivv}.

In this work, we consider apparent horizon cutoff in non-flat HDE cosmology. Matter contents are NMDC scalar field, holographic vacuum energy and dust matter. We study dynamical effects of HDE, NMDC field and spatial curvature with kinematical implication of late-time expansion. We perform dynamical analysis to the system and we compare our results to the observed expansion history obtained from SNIa only schematically.   
 Recent work by \cite{Sushkov:2023aya} considers qualitatively cosmological dynamics of NMDC gravity for non-flat universe without holographic effects. The work agrees with ours in non-holographic limit. In addition, we show stability analysis of fixed points of the model.    Indeed, the sign of the coupling $\k$ in equation (\ref{NMDCaction}) could be opposite to the sign of the $\varepsilon g_{\mu\nu}$ term. One can see that phantom effect with $\varepsilon = -1$ is not the only negative kinetic energy contribution, but also the NMDC term. Hence, for $\varepsilon = 1$, the negative kinetic energy contribution can as well come from the NMDC sector. We introduce the field equation in section \ref{secfieldequation}. Considerations of flat case and non-flat case are of section \ref{secflat} and \ref{secnonflat} where idea of using gravitational constant in holographic vacuum density is discussed. Dynamical stability analysis is shown in section \ref{secnonflat} and numerical integration kinematic results on Hubble parameter and equation of state parameter are presented in section \ref{secnumer}. At last, conclusion and discussion are in section \ref{seccon}.


\section{Holographic dark energy with NMDC gravity effect}  \label{secfieldequation}
The field equation derived from the NMDC action (\ref{NMDCaction}) is
\begin{equation}  \label{fieldeq}
	G_{\mu\nu} = 8\pi G \l( T^{(\rm {m})}_{\;\mu\nu}+T^{(\phi)}_{\;\mu\nu}+\kappa\Theta_{\mu\nu} \r)  - \Lambda g_{\mu\nu}\,.
\end{equation}
 The stress tensor of dust matter  is  $T_{\mu\nu}^{(\rm m)}$. The scalar field stress tensor, $T^{(\phi)}_{\;\mu\nu}$,  and the NMDC stress tensor, $\Theta_{\mu\nu}$, read
\begin{eqnarray}
	T^{(\phi)}_{\;\mu\nu} & = &  \varepsilon (\nabla_{\mu}\phi) (\nabla_{\nu}\phi) - \frac{\varepsilon}{2}g_{\mu\nu}(\nabla \phi)^2 - g_{\mu\nu}V(\phi), \\
	\Theta_{\mu\nu}  &=& -\frac{1}{2}(\nabla_{\mu}\phi)(\nabla_{\nu}\phi) R + 2 (\nabla_{\alpha}\phi) \nabla_{\left(\mu \right.}\phi R^{\alpha}_{\left.\;\nu\right)}   + (\nabla^{\alpha}\phi) (\nabla^{\beta}\phi) R_{\mu\alpha\nu\beta}
	+ (\nabla_{\mu}\nabla^{\alpha}\phi) (\nabla_{\nu}\nabla_{\alpha}\phi)  \\
	&& - (\nabla_{\mu}\nabla_{\nu}\phi) \Box\phi - \frac{1}{2}(\nabla\phi)^2 G_{\mu\nu} 
	  +g_{\mu\nu}\left[-\frac{1}{2}(\nabla^{\alpha}\nabla^{\beta}\phi)(\nabla_{\alpha}\nabla_{\beta}\phi) + \frac{1}{2}(\Box\phi)^2  
	  - (\nabla_{\alpha}\phi)(\nabla_{\beta}\phi) R^{\alpha\beta} \right].
\end{eqnarray}
There are conservations
\bea
	[\varepsilon g^{\mu\nu}+\kappa G^{\mu\nu}]\nabla_{\mu}\nabla_{\nu}\phi &=& -V_{\phi},  \label{kggg}  \\
	\nabla^{\mu}[T^{(\phi)}_{\;\mu\nu}+\kappa \Theta_{\mu\nu}]  &=& 0,
\eea
where $V_{\phi}\equiv \d V(\phi)/\rm{d}\phi$. These conservations are consequence of the Bianchi identity $\nabla^{\mu} G_{\mu\nu}=0$ and the conservation of matter field $\nabla^{\mu}T^{({\rm m})}_{\;\mu\nu}=0$. 
 Dust matter density is denoted as $\rho_{\rm m}$. Quantum gravity motivates phenomenological  
 holographic energy scale cutoff to the vacuum energy density, implying that it could not exceed  
 $\rho_{\Lambda} = \Lambda/(8\pi G) = {3c^2}/{(8\pi G L^2)}$. The IR cutoff length scale, $L$ motivated from quantum gravity effect, is introduced, not in the classical Lagrangian, but in the vacuum density. Other separated conservation equations are
  $\dot{\rho}_{\Lambda} +3H(\rho_{\Lambda} + P_{\Lambda})=0$  and  $\dot{\rho}_{\rm m} = -3H\rho_{\rm m}$
where $P_{\Lambda} $ is pressure of the holographic vacuum energy.

\section{Flat case}\label{secflat}
In flat case, without the holographic effect, the NMDC Friedmann equation can be viewed in two ways, 
\begin{eqnarray}
	H^2 &=&\frac{8\pi G}{3}\left[\frac{1}{2}\dot{\phi}^2(\ve-9\kappa H^2)+V(\phi)+\rho_{\Lambda}+\rho_\m\right],     \label{F1}   \eea  or
	\bea
	H^2 & = & \frac{8\pi G_{\text{eff}}}{3}\left[\frac{\ve }{2}\dot{\phi}^2+V(\phi)+\rho_{\Lambda}+\rho_\m \right].   \label{F2}
\end{eqnarray}
 Both equations are the same, however it is possible to interpret them in two different pictures, i.e. either modification of the kinetic term of the scalar field, $(1/2)\dot{\phi}^2(\varepsilon-9\kappa H^2)$ or modification of the gravitational constant. 
  The equation (\ref{F1}) can be viewed as a flat FLRW universe evolving with conventional gravitational constant, $G$, and the universe is filled with matter field, vacuum energy and the NMDC (or phantom NMDC) field while the equation (\ref{F2}) represents a universe with effective gravitational constant $G_{\rm eff}$, 
\begin{eqnarray} \label{5}
	G_{\text{eff} }(\dot{\phi}) \; \equiv \;  \frac{G}{1+12\pi G \k \dot{\phi}^2}, 
\end{eqnarray}
 and the universe is filled with matter field, vacuum energy and canonical (or phantom) field. The conservation of the NMDC field is a result of equation (\ref{kggg}), the Klein-Gordon equation, 
 \begin{equation}\label{kg}
 	\ddot{\phi}+3H \dot{\phi}=-\frac{V_\phi}{\varepsilon - 3\kappa H^2} + \frac{6 \kappa H \dot{H} \dot{\phi}}{\varepsilon - 3\kappa H^2}.
 \end{equation}
 From now on, we consider only $\ve = 1$ which is non-phantom case. 
 These give us two choices of gravitational constant in making definitions of the vacuum energy density (equation (\ref{1})).  The vacuum energy density should read either
 $\rho_{\Lambda} = {3c^2}/{(8\pi G L^2)}$ or $\rho_{\Lambda} = {3c^2}/{(8\pi G_{\rm eff} L^2)}$ according to which pictures we interpret. 
 Apparent horizon cutoff length scale for flat case is the Hubble length,   i.e.  $L = H^{-1}$.  Hence the holographic vacuum energy are either
 \begin{equation}\label{rhol}
 	\rho_{\Lambda}=\frac{3c^2H^2}{8\pi G}\,, 
 	\end{equation}
 		 or \begin{equation}\label{rhol2} 	\rho_{\Lambda}=\frac{3c^2H^2}{8\pi G_{\text{eff}}}=\frac{3c^2 H^2}{8 \pi G}(1+12\pi \kappa G \dot{\phi}^2).
 \end{equation}
 The second choice, i.e. choosing equations (\ref{F2}) and (\ref{rhol2}), has been shown in 
 \cite{Baisri:2022ivv}  to have shortcomings in giving consistent inflationary parameters and, moreover at late time, when constrained with variation rate of gravitational constant, it favors $\k > 0$. At non-holographic limit, this is in conflict with the results given in \cite{Tsujikawa:2012mk} of which $\k < 0 $ is required for inflation. Note that notation in \cite{Tsujikawa:2012mk} differs from ours. Therefore, for the flat case, we should restrict our consideration to the  first choice. That is equation (\ref{rhol}), $\rho_{\Lambda}= {3c^2H^2}/{8\pi G}$, with the Friedmann equation (\ref{F1}). We will see in the next sections if there is any modification to its dynamics when the space is curved.

 \section{Non-flat case}  \label{secnonflat}
 Non-zero curvature case allows richer characters of the holographic NMDC cosmology. With non-zero $k$, cosmological dynamical behavior could be modified with the curvature terms. The apparent horizon, 
\be
R_{\rm A} =  \f{1}{\sqrt{H^2 + k/a^2}}\,,
\ee 
 reduces to Hubble length when taking $k = 0$. For the sake of analogy to the flat case, we should consider the holographic vacuum density in two cases. These are $\rho_{\Lambda}= {3c^2H^2}/{8\pi G R_{\rm A}^2}$ and $\rho_{\Lambda}= {3c^2}/{8\pi G_{\rm eff} R_{\rm A}^2}$.  The field equation (\ref{fieldeq}) gives a modified Friedmann equation,
 \begin{equation}    \label{fff1}
 	H^2+\frac{k}{a^2}\:=\:
 	\f{8\pi G}{3}\left\{
 	\frac{\dot{\phi}^2}{2}
 	\left[
 	1-\kappa
 	\left(
 	9 H^2+\frac{3 k}{a^2}
 	\right)
 	\right]
 	+V(\phi)+\rho_{\rm m}+\rho_\Lambda
 	\right\}\,,
 \end{equation} 
 which, in similar spirit to $G_\text{eff}(\dot{\phi})$ in the flat case (equation (\ref{F2})), is expressed as
 \begin{equation}   \label{f99}
 	3H^2+\frac{3k}{a^2}=8\pi G_{\text{eff}}
 	\left[
 	\frac{\dot{\phi}^2}{2}+V(\phi)+\frac{3\kappa k\dot{\phi}^2}{a^2}+\rho_{\rm m}+\rho_\Lambda
 	\right],
 \end{equation} 
 where $
 	G_{\text{eff}}(\dot{\phi}) \equiv {G}/({1+12\pi G\kappa\dot{\phi}^2})$. Since there is a NMDC-curvature coupling term, ${3\kappa k\dot{\phi}^2}/{a^2}$ in equation (\ref{f99}), one can see that the MMDC character cannot be fully extracted into  the $G_{\text{eff}}(\dot{\phi})$ term. This is unlike the flat case ($k=0$) (equation \eqref{F2}) of which the NMDC effect is fully incorporated in the $G_{\text{eff}}$. It is known that for the NMDC theory, one cannot express effective gravitational constant at the Lagrangian level. For the flat case, the effective gravitational constant may be written at the Friedmann equation level.  Since  consideration of non-zero curvature is more generic, we conclude that effective gravitational constant cannot be realized at the Friedmann equation level.  As a result, using  $G_{\text{eff}}$ in vacuum energy density as $\rho_{\Lambda}= {3c^2}/{8\pi G_{\rm eff} R_{\rm A}^2}$ is not plausible.   
 We shall therefore consider only the $\rho_{\Lambda}= {3c^2}/{8\pi G R_{\rm A}^2}$ case.
 The Friedmann  equation (\ref{fff1}) and the equation,
\begin{equation}
	2\dot{H}+3H^2+\frac{k}{a^2}
	\:=\: -8\pi G
	\left\{
	\frac{\dot{\phi}^2}{2}   \left[ 1 +   \kappa 
	\left(
	2\dot{H}+3H^2+4H\ddot{\phi}\dot{\phi}^{-1}-\frac{k}{a^2}
	\right) \right]
	-V(\phi)+P_{\rm m}+P_\Lambda
	\right\}\,,     \label{Ray}
\end{equation}
are derived from the field equation (\ref{fieldeq}) and the non-flat Klein-Gordon equation is derived from equation (\ref{kggg}),
\begin{equation}
	\ddot{\phi}+3H\dot{\phi}\:=\:\frac{-V_{\phi}+6\kappa H\dot{H}\dot{\phi}-6\kappa H\dot{\phi} {k}/{a^2}}{1-3\kappa(H^2+ {k}/{a^2})},	
\end{equation}
which can be rewritten as,
\be
	\ddot{\phi} \l[ 1 - 3 \k \l(   H^2  + \f{k}{a^2}    \r)      \r]     +  3H\dot{\phi} \l[  1 -  \k \l(  2  \dot{H}  + 3 H^2 + \f{k}{a^2}  \r)      \r]   \:=  \:	- V_{\phi}\,.
\ee
As seen in the above field equations, this is the FLRW universe with gravitational constant $G$. The scalar kinetic term ($\dot{\phi}$ term) and scalar dynamical term ($\ddot{\phi}$ term) are all modified with the NMDC coupling.     
We notice that non-minimal derivative coupling to gravity ($\k$ term) does not only couple to only the kinematic sector, i.e. to $H$ or $\dot{H}$ but also couples to spatial curvature $k$. Hence the spatial curvature could have some effects to the dynamics. In this case, the holographic vacuum energy density is  
 \begin{equation}
 	\rho_\Lambda \: = \:  \f{3c^2H^2}{8\pi G R_{\rm A}^2} \: =   \:        \frac{3c^2}{8\pi G}\left(H^2+\frac{k}{a^2}\right).    \label{rhoah}
 \end{equation}
Considering power-law potential, $V(\phi) = V_0 \phi^n $ for $V_0 \geq 0$ and $n$ is an even positive number, we define dimensionless dynamical variables as (see e.g. \cite{Roy:2014yta} for quintessence case),
 \begin{equation}  \label{dimless_var}
 	\begin{split}  
 		x& \equiv \frac{8\pi G\dot{\phi}^2}{6H^2},\:\;\;\; y \equiv \frac{8\pi G V_0 \phi^n}{3H^2},\:\;\;\; r \equiv -12\pi G\kappa\dot{\phi}^2,\: \;\;\; s \equiv -\frac{4\pi G\kappa k\dot{\phi}^2}{a^2H^2},
 		\\ 
 		\Omega_{\rm m}& \equiv \frac{8\pi G\rho_{\rm m}}{3H^2}, \:\;\;\; \Omega_\Lambda  \equiv c^2\left(1+\frac{k}{a^2H^2}\right),\: \;\;\; \Omega_k  \equiv -\frac{k}{a^2H^2}\,,
 	\end{split}
 \end{equation}
such that the Friedmann equation (\ref{fff1}) is written as  
\begin{equation}
	1=x+y+r+s+\Omega_{\rm m}+\Omega_\Lambda+\Omega_k.
\end{equation}
These dimensionless variables are not independent. Some of these variables can be expressed in terms of the others, i.e. 
\begin{equation}
	\Omega_\Lambda 
	=c^2 \l(1-\Omega_k \r), \;\;\;\;
	  \text{and}  \;\;\;\;  	s = -\frac{r\Omega_k}{3}.  \label{omalamd_omegak}
\end{equation}
Hence the Friedmann constraint becomes 
\begin{equation}
	1\;=\;x+y+r-\frac{r\Omega_k}{3}+\Omega_{\rm m}+c^2(1-\Omega_k)+\Omega_k,
\end{equation}
or
\begin{equation}
	\Omega_{\rm m} \;=\; 1-x-y-r+\frac{r\Omega_k}{3}-c^2(1-\Omega_k)-\Omega_k.
	\label{friedmann}
\end{equation}
We can express $\Omega_m$ in terms of $x, y, r, \Omega_k$.  
Autonomous system of these variables is therefore 
\begin{equation}
	\begin{split}
		x'&\;=\; 2x\left(\epsilon-\delta\right),\\
		y'&\;= \;2y\left(\frac{1}{2}nu+\epsilon\right),\\
		r'&\;=  \;-2 r \delta,\\
		u'&\;= \; u(\epsilon-\delta-u),	\\
		\Omega_k'&\;=\; 2\Omega_k(\epsilon-1),
	\end{split} \label{auto}
\end{equation}
where we define
\begin{equation}
	\delta=-\frac{\ddot{\phi}}{H\dot{\phi}}\,,\;\;\;\epsilon=-\frac{\dot{H}}{H^2}\,, \;\;\;
\text{and}\;\;\; 
 u=\frac{\dot{\phi}}{H\phi}\,\,.
\end{equation}
According to the field equations,	
$
		3H^2+ {3k}/{a^2} =  8\pi G\rho_\text{tot}  \;  \text{and}  \;
		2\dot{H}+3H^2+ {k}/{a^2} =  -8\pi G P_\text{tot}	
$
the effective equation of state coefficient, $ w_\text{eff} = {P_\text{tot}}/{\rho_\text{tot}} $, is hence
\begin{equation}
	w_\text{eff}=\frac{-1 + ({2\epsilon}/{3}) + ({\Omega_k}/{3})}{1-\Omega_k}\,.  \label{weffep}
\end{equation}
The autonomous system is closed because $\delta$ and $\epsilon$ can be expressed in term of the other dynamical variables, i.e.
\begin{equation}
	\begin{split}     \label{eq:epsilon}
		\epsilon\;=\;\Big\{&-3r\left[c^2\left(\Omega_k^2-4\Omega_k+3\right)-2nuy+4x (\Omega_k-3)-3y\Omega_k+3y-\Omega_k^2+4\Omega_k-3\right]
	\\
	&+9x\left[c^2 (\Omega_k-3)+3x-3y-\Omega_k+3\right]+r^2\left(\Omega_k^2-2\Omega_k+9\right)\Big\}\Big/
	\\
	&
	\Big\{-18x(c^2-1)+2r^2(3+\Omega_k)+6r \l[c^2(\Omega_k-1)-x-\Omega_k+1 \r]\Big\},
\end{split}
\end{equation}
and
\begin{equation}
	\begin{split} \label{eq:delta}
			\delta=\frac{-18 x \left(3 c^2+2 r-3\right)+18 r y+4 r^2\Omega_k-3 n u y \left(3 c^2+r-3\right)}{-18x(c^2-1)+2r^2(3+\Omega_k)+6r[c^2(\Omega_k-1)-x-\Omega_k+1]}.
	\end{split}
\end{equation}
According to \cite{Skugoreva:2013ooa}, there exists a relation among  the variables $u, r, x$ and $y$: 
\begin{equation}
	ryu^n+3\kappa V_06^{\frac{n}{2}}(8\pi G)^{\frac{2-n}{2}}x^{\frac{n+2}{2}}\;=\;0.
	\label{const}
\end{equation}
Hence, there exists another constraint $u=u(x,y,r)$ where $n, \kappa, V_0, c$ are numerical parameters. We shall solve the autonomous system straightforwardly, and exclude solutions that do not satisfy the constraint  \eqref{const}. Fixed points are to be found as the system (\ref{auto}) is set to zero. Stability of the fixed points is found considering linear perturbation, 
$
		x=x_c+\delta x,\; y=y_c+\delta y,\; r=r_c+\delta r,\;  u=u_c+\delta u,\;  \text{and} \;
		  \Omega_k=\Omega_{kc}+\delta \Omega_k
$ in the autonomous system (\ref{auto}) where subscription $c$ denotes the fixed points.
Linearizing the autonomous system, the first order perturbation can be expressed as 
\begin{equation}
	\frac{{\rm d}}{{\rm d}N}
	\begin{pmatrix}
		\delta x\\
		\delta y\\
		\delta r\\
		\delta u\\
		\delta\Omega_k
	\end{pmatrix}
	=
	\mathcal{M}
	\begin{pmatrix}
		\delta x\\
		\delta y\\
		\delta r\\
		\delta u\\
		\delta\Omega_k
	\end{pmatrix}, 
\end{equation}
where Jacobian matrix $\mathcal{M}$ can be defined as  
\begin{equation}
	\mathcal{M}=\begin{pmatrix}
		\partial_x x' & \partial_y x' & \partial_rx'  &\partial_ux' &\partial_{\Omega_k}x'
		\\
		\partial_x y' & \partial_y y' & \partial_ry'  &\partial_uy'  &\partial_{\Omega_k}y'
		\\
		\partial_xr' & \partial_yr' & \partial_rr'  &\partial_ur'  &\partial_{\Omega_k}r'
		\\
		\partial_xu' & \partial_yu' & \partial_ru'  &\partial_uu'  &\partial_{\Omega_k}u'
		\\
		\partial_x\Omega_k' & \partial_y\Omega_k' & \partial_r\Omega_k'  &\partial_u\Omega_k' &\partial_{\Omega_k}\Omega_k'
		\\
	\end{pmatrix}_\text{at\;fixed\;points}
\end{equation}
and $\partial_x x', \partial_y x', \partial_r x', \partial_u x', \partial_{\Omega_k} x'$ denote differentiation of $x'$ with respect to $x,y,r,u$ and $\Omega_k$ respectively.
Performing linear stability analysis, eigenvalues of the Jacobian matrix can identify stabilities of the fixed points. 
The Jacobian is $5 \times 5$ matrix, hence there are five eigenvalues.  
A fixed point is asymptotically stable if all eigenvalues are negative. It is unstable if all eigenvalues are positive. A fixed point is saddle point if at least one eigenvalue is positive. Linear stability theory fails to determine stability when all eigenvalues are zero or when some are zero and some are negative. In this case, we use numerical integration result to determine stability of the fixed point. Characters of fixed points are shown in table \ref{tablexx1}.

\begin{table}[h]
	\centering
	\renewcommand{\arraystretch}{1.5} 
	\begin{tabular}{|c|c|c|c|c|c|c|c|c|c|c|c|} 
		\hline\hline
		\multirow{2}{*}{{Names}}  &  \multicolumn{8}{|c|}{{Fixed points}} & 
		\multirow{2}{*}{{$w_{\text{eff}}$}} & \multirow{2}{*}{{stability}}& \multirow{2}{*}{{existence}}\\  
		\cline{2-9}
		& $x_c$ &  $y_c$ &  $r_c$& $\;\;s_c\;\;$ & $\;\;u_c\;\;$ & $\;\;\Omega_{kc}\;\;$ & $\;\;\Omega_{\Lambda c}\;\;$ & $\Omega_{{\rm m}c}$ &&&
		\\
		\hline\hline
		(a) & 0  & 0 & $r$ & 0 &0 & 0 & $c^2$  & $1-c^2-r$ & $0$ 
		& unstable & 
		case (1): $r = 0$ with $\k \neq 0,$    
		\\
		&&&&&&&&&&& case (2):  $0 < r \leq 1-c^2$ with $\kappa<0$
		\\
			\hline
		(b) & 0 & $1-c^2$ & 0 & 0& 0  & 0 & $c^2$ & 0 & $-1$ 
		& stable &        for  $\forall \kappa$                              
		\\
			\hline 
		(c) & $\frac{-1+c^2}{2}$ & 0 & $\frac{3(1-c^2)}{2}$ &0 &0  & 0 & $c^2$ & 0 & $-1$ 
		& saddle & for $ \k <0 $ 
		\\
			\hline
		(d) & $1-c^2$ & 0 & 0& 0 & 0 & 0 & $c^2$ & 0 & 1 & saddle & $V_0=0$ and $\k = 0$
		\\
		\hline   
	\end{tabular}  
\caption{Fixed points expressed in all dynamical variables and their stabilities  \label{tablexx1} } 
\end{table}

\subsubsection{Fixed Point {\rm (a)}} 
In this case the eigenvalues reads 
\begin{equation}
	\mu_1=0,\;\; \mu_2=1,\;\; \mu_3=\frac{3}{2},\;\; \mu_4=3,\;\; \mu_5=3.
\end{equation}  
Since one eigenvalue is zero and others are positive, this point represents an unstable node for all $n$ and $\kappa$. The dynamical parameter $r$ is arbitrary in range $0\leq r \leq 1-c^2$ whereas the parameter $c$ ranges within $0\leq c < 1$. Substituting fixed points into constraint equation \eqref{omalamd_omegak} and \eqref{friedmann}, we find that density parameter of matter and holographic effect are $\Omega_{{\rm m}c}=1-c^2-r$ and $\Omega_{\Lambda c}=c^2$ respectively. The point corresponds to $w_{\text{eff}}=0$. 
Substituting the fixed point coordinate into equation (\ref{eq:epsilon}), we find $\epsilon= {3}/{2}$. With $\epsilon=-\dot{H}/H^2$, the fixed point corresponds to dust-dominated  solution,
\begin{equation}\label{eq:scal fact fp1}
	H(t)=\frac{2}{3(t-t_0)}\,,    \;\;\;\; \text{or} \;\;\;\;
	a(t)=a_0(t-t_0)^{2/3},
\end{equation} 
where $a_0$ and $t_0$ are an initial value of scale factor and an initial value of time respectively. 
Integrating $r=r_c=-12\pi G\kappa\dot{\phi}^2$, gives a solution 
\begin{equation}\label{eq:scal field fp1}
	\phi(t)=\sqrt{\frac{r}{-12\pi G\kappa}}(t-t_0)+\phi_0,
\end{equation}
where $\phi_0$ is some initial values. These solutions do not have any holographic effects. For the scalar field to be real, it is either case (1): $r=0$ (i.e. $\dot{\phi} = 0$) for all real value of $\kappa$ except $\kappa = 0$ or case (2): $0 < r \leq 1-c^2$ with $\kappa<0$. Substituting these solutions into dimensionless variables in equation \eqref{dimless_var}, for $r=0$ and for all real value of  $\kappa$, we have constant field solution, $\phi=\phi_0$.  For $0 < r \leq 1-c^2$ and $\kappa<0$, then $\phi \propto t $, recovering the NMDC result reported earlier \cite{Skugoreva:2013ooa}.
The point (a) is the effective matter-dominated case where there are two components, $\Omega_{\rm m}$ and $\Omega_{\Lambda}$ driving evolution of the universe. If there is no holographic effect, the point is purely matter-dominating fixed point, $\Omega_{{\rm m} c}=1$.  If there is only holographic component, without any other matter in flat universe, $c=1$ is allowed and this is not a singularity. This case corresponds to  
 $\Omega_{\Lambda c}=1$ which gives dust-like evolution as mentioned by Hsu in \cite{Hsu}.


\subsubsection{Fixed Point {\rm (b)}}
\begin{figure}
  	\centering
  	\includegraphics[width=5.2cm]{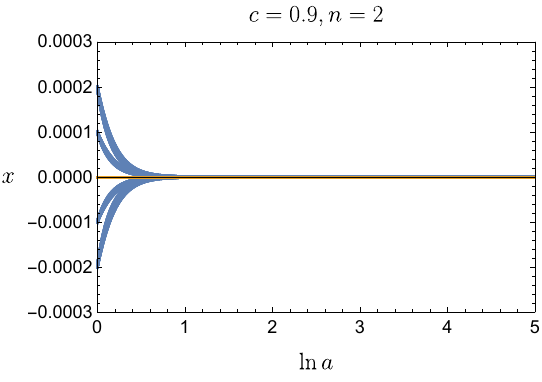}
  	\includegraphics[width=5.0cm]{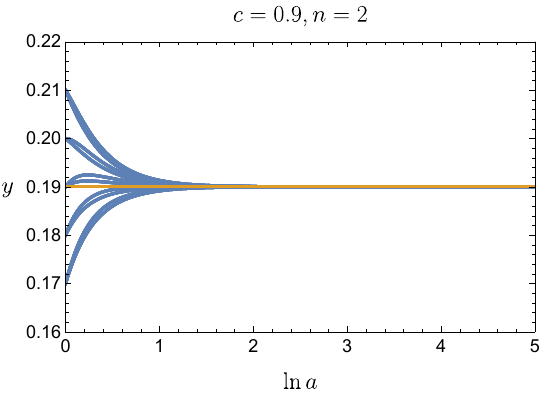}
  	\includegraphics[width=5.2cm]{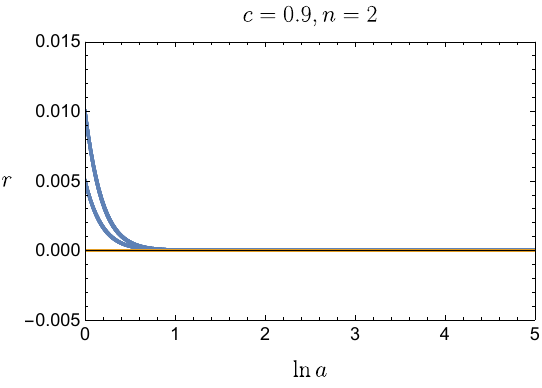}
  	\\
  	\includegraphics[width=5.3cm]{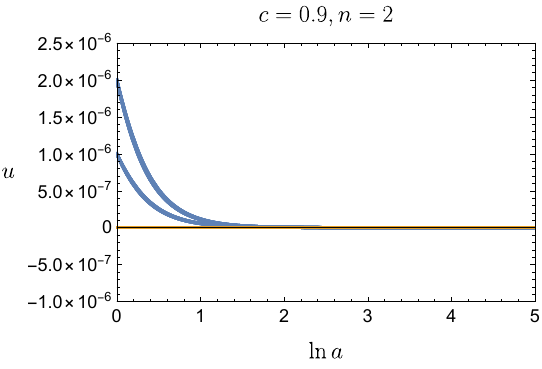}
  	\includegraphics[width=5.3cm]{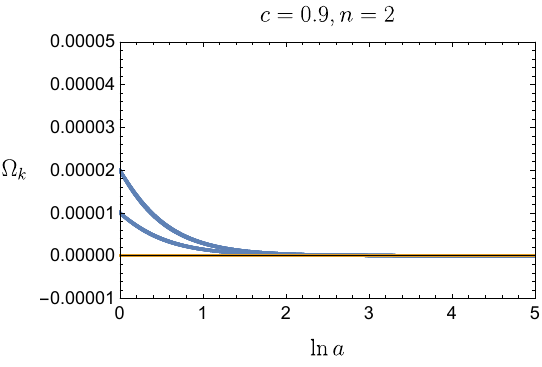}
  	\caption{Numerical integration of the autonomous system with respect to $\ln a$ around the fixed point (b) where $n=2$ and $c=0.9$. 
  		The orange lines mark fixed point value. The blue lines are numerical solutions with different initial conditions.}
  	\label{plotfigev}
  \end{figure}

The eigenvalues are 
\begin{equation}
	\mu_1=0,\;\; \mu_2=0,\;\; \mu_3=-3,\;\; \mu_4=-3,\;\; \mu_5=-2\,.
\end{equation}  
Since the eigenvalues are zero and negative, this point is non-hyperbolic and the linear stability analysis fails to identify character of the fixed point. The point exists for all $n$ with equation of state corresponds to that of cosmological constant, i.e. $w_\text{eff}=-1$.
 With the constraint equations \eqref{omalamd_omegak} and \eqref{friedmann}, we have $\Omega_{{\rm m}c}=0$ and $\Omega_{\Lambda c}=c^2$ respectively. At this point, the potential and holographic effect play a role of cosmological constant solution as $\Omega_k \rightarrow 0$ (asymtotically flat) at late time, $t \rightarrow \infty$.   The fixed point and the constraints \eqref{omalamd_omegak} and \eqref{friedmann} imply $\Omega_{{\rm m}c}=0$ and $\Omega_{\Lambda c}=c^2$. With the equation \eqref{eq:epsilon}, $\epsilon=0$. From $\epsilon= -\dot{H}/H^2$, 
  \begin{equation}\label{eq:scal fact fp2}
  	H= \pm \sqrt{\lambda_1}, \quad \text{or} \quad	a(t)=a_0e^{\pm \sqrt{\lambda_1} t},
  \end{equation} 
  where $\lambda_1 > 0$ is constant.
  Using definition of $y$, the scalar solution is a constant function, i.e.
  \begin{equation}\label{eq:scal field fp2}
  	\phi=\left( \frac{3\lambda_1 (1-c^2)}{8\pi GV_0}\right)^{\frac{1}{n}}=\phi_0\,,
  \end{equation} 
  where $
  	\lambda_1 \equiv {8\pi G V_0\phi_0^n}/\l[{3(1-c^2)}\r]\,.
  $
  Using the solutions \eqref{eq:scal fact fp2} and \eqref{eq:scal field fp2} in \eqref{dimless_var}, we can see qualitatively that, $x\rightarrow x_c=0,\, y\rightarrow y_c=1-c^2,\, r\rightarrow r_c=0,\, u\rightarrow u_c=0,\, \Omega_k\rightarrow \Omega_{kc}=0$ as $t\rightarrow\infty$  for any real value of $n$ and for any real value of $\kappa$ with $V_0>0$ and $0 \leq c < 1$. 
  Moreover,  given small numerical perturbation of initial condition around the fixed point in numerical integration result, evolution of the autonomous system is presented in figure \ref{plotfigev}. This points out that the fixed point (b) is a stable node.

\subsubsection{Fixed Point {\rm (c)}} 
\begin{figure}[h]
	\centering
	\includegraphics[width=5.2cm]{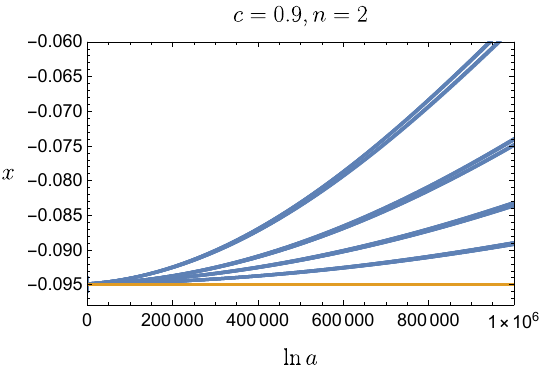}
	\includegraphics[width=5cm]{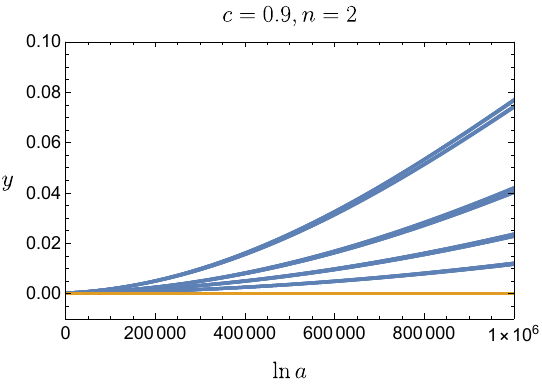}
	\includegraphics[width=5cm]{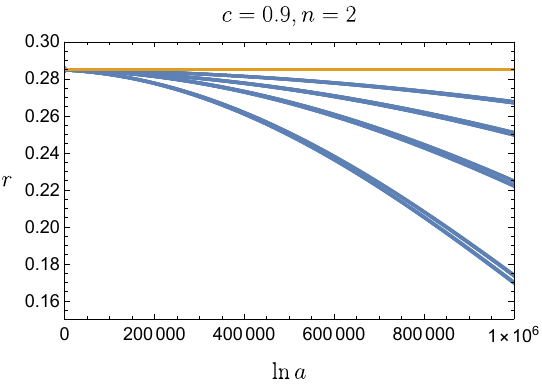}
	\\
	\includegraphics[width=5.2cm]{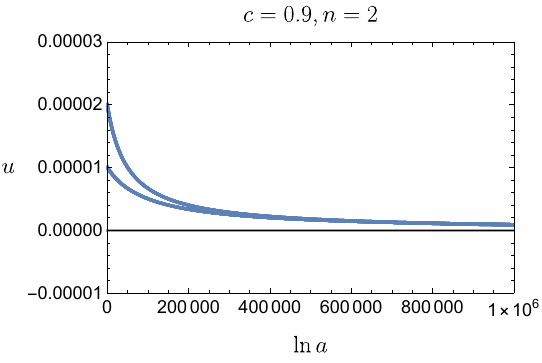}
	\includegraphics[width=5.2cm]{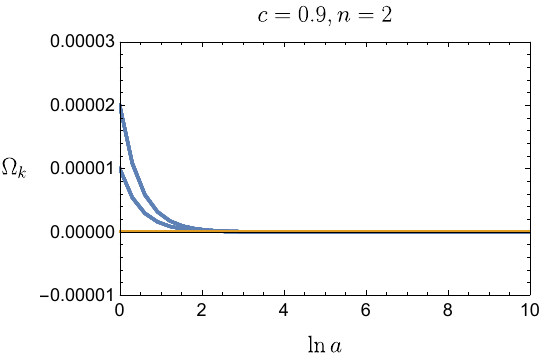}
	\caption{Numerical integration of the autonomous system with respect to $\ln a$ around the fixed points (c) where $n=2$ and $c=0.9$, the orange lines represent the fixed point and the blue curves are the numerical solutions.}  \label{fff2} 
\end{figure}

This point has eigenvalues, 
\begin{equation}
	\mu_1=-3,\;\; \mu_2=-3,\;\; \mu_3=-2,\;\; \mu_4=0,\;\; \mu_5=0\,,
\end{equation}  
which are zero and negative. It is also a 
non-hyperbolic point as similar to the point (b) and the linear stability fails to tell character of the point. 
We will use numerical integration to check its stability as we did for the point (b). 
At the fixed point (c), the equations \eqref{eq:epsilon} and \eqref{eq:delta} are $\epsilon=0$ and $\delta=0$.  From equation \eqref{weffep},
the point (c) corresponds to $w_\text{eff}=-1$.  Since NMDC field could be a cause of acceleration, we should find scalar field solution before finding the effect which is the corresponding scale factor function. From equation \eqref{dimless_var} and the table \ref{tablexx1}, we see that $x \equiv 8 \pi G \dot{\phi}^2/(6H^2)$ and $x_c = (-1 + c^2)/2$ as well as $r \equiv -12 \pi G \k \dot{\phi}^2$ and $r_c = 3(1-c^2)/2$. At $x = x_c$ 
(with $H^2 = (3 \k)^{-1}$ to be found later) and $r = r_c$, one can find scalar field solution, 
\begin{equation}\label{eq:scal field fp3}
	\phi(t)= \sqrt{-\frac{(1-c^2)}{8\pi G\kappa}}(t-t_0)+\phi_0\,,
\end{equation}
which is real if $\k < 0$.  At the point (c), integrating the relation, $\epsilon=-\dot{H}/H^2 = 0$, we obtain a solution,
\begin{equation}\label{eq:scal fact fp3}
	H = \pm \sqrt{\lambda_2}, \quad \text{or}      \quad         	a(t)=a_0 e^{\pm \sqrt{\lambda_2}t},
\end{equation} 
where $\lambda_2$ is a constant. At the fixed point (c),  ${x_c}/{r_c}=-{1}/{3}$. 
Using the definition of $x$ and $r$ in \eqref{dimless_var}, we have $x/r   = -(9\k H^2)^{-1}$ (where $x=x_c$ and $r=r_c$), leading to $H^2 = (3 \k)^{-1}$. 
Since we have $\k < 0$, therefore $\lambda_2 = (3 \k)^{-1} < 0$  and $\sqrt{\lambda_2}$ is imaginary. 
We write $\sqrt{\lambda_2} = i \sqrt{|\lambda_2|} = i/\sqrt{3 |\k|}$, therefore
$
	H = \pm {i}/{\sqrt{3 |\k|}}$, or  $ a(t)=a_0 \exp[{\pm i  {t}/{\sqrt{3 |\k|}}  }].
 $
This corresponds to an oscillating solution, i.e.
$
	a(t)=a_0 [\cos (  {t}/{\sqrt{3| \kappa |}}) \pm i \sin (  {t}/{\sqrt{|\k|}} )].
$
Taking only real part of the solution, the Hubble parameter reads, 
$
	H(t) = -{1}/{ \sqrt{3|\kappa|} } \tan (  {t}/{\sqrt{3|\k|} } )\,.
$
Numerical integration is performed as small perturbation (from the fixed point (c)) is introduced to the system. The results are presented in figure \ref{fff2} manifesting divergent and convergent evolution of the dimensionless parameters away from the fixed point and to the fixed point (c). Therefore the point (c) is a saddle node.

\subsubsection{Fixed Point {\rm (d)}} 
In this case, the eigenvalues read, 
\begin{equation}
	\mu_1=-6,\;\; \mu_2=6,\;\; \mu_3=4,\;\; \mu_4=3,\;\; \mu_5=0.
\end{equation}  
This point represents a saddle point because the eigenvalues are mixed positive and negative.  The kinetic term and holographic vacuum density term are dominant. 
At this point, we obtain $\epsilon=-\dot{H}/H^2=3$, corresponding to stiff fluid with $w_\text{eff}=1$. This gives 
\begin{equation}\label{eq:scal fact fp4}
H(t)=\f{1}{3(t-t_0)}                   , \quad \text{or}      \quad                                       	a(t)=a_0(t-t_0)^{1/3}\,.
\end{equation}
The fixed point coordinate, $x_c =1-c^2$, with definition in equation \eqref{dimless_var} implies  
\begin{equation}\label{eq:scal field fp4}
	\phi(t)=\sqrt{\frac{1-c^2}{12\pi G}}\ln(t-t_0)+\phi_0\,.
\end{equation}
At $y_c=0$, with $\phi(t)$ and $H(t)$ solutions \eqref{eq:scal fact fp4} and \eqref{eq:scal field fp4}, we have
$y_c = 8 \pi G V_0 \phi^n (t - t_0) = 0$ which is valid only when $V_0 = 0$.   
The fixed point condition $r_c = 0 = -12\pi G \k \dot{\phi}^2$ is valid only when $\k = 0$.

\section{Numerical Solutions}  \label{secnumer} 
\subsection{Flat Case}
\begin{figure}[h]
	\centering
	\includegraphics[width=15.5cm]{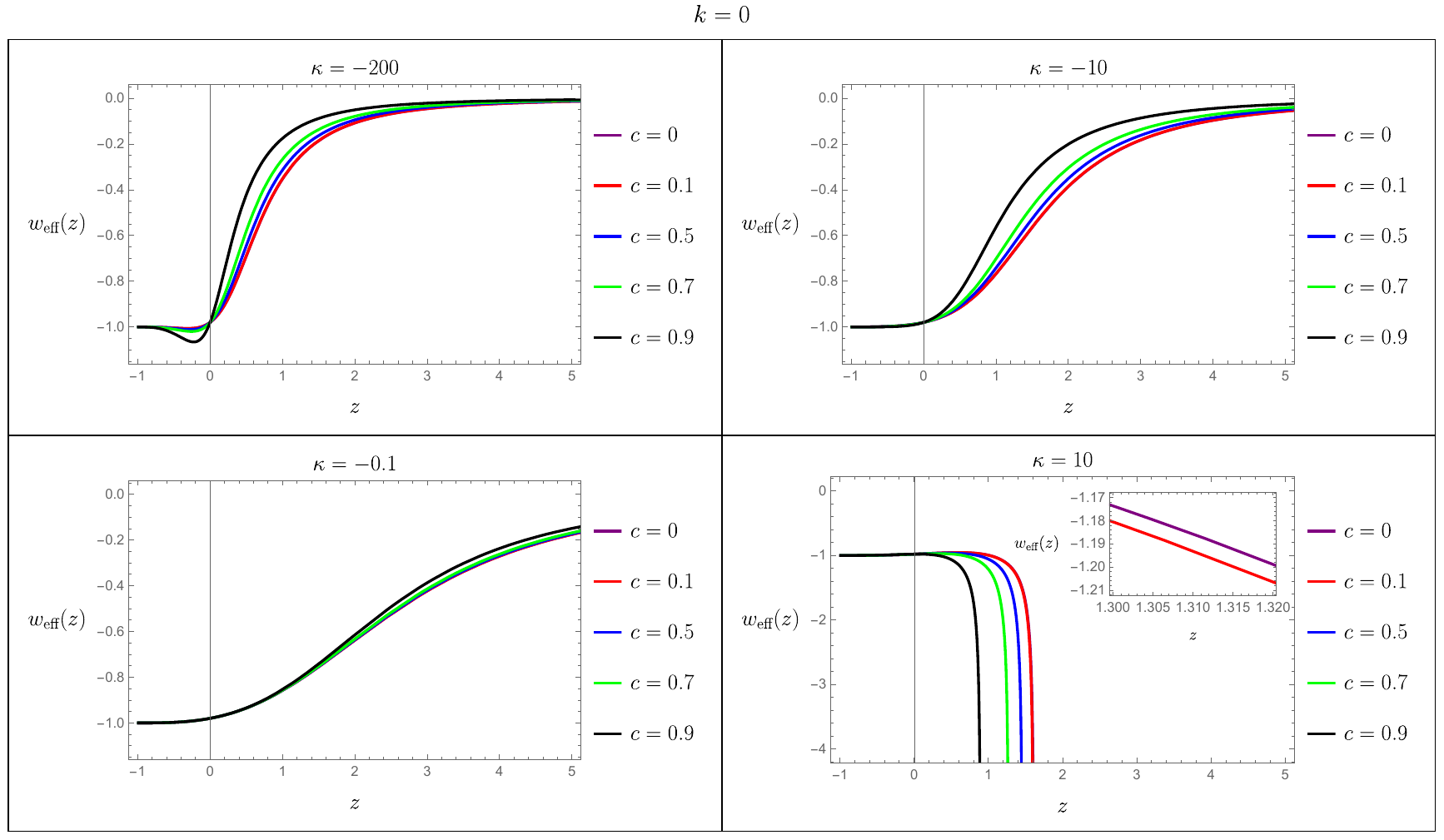}
	\centering
	\includegraphics[width=15.5cm]{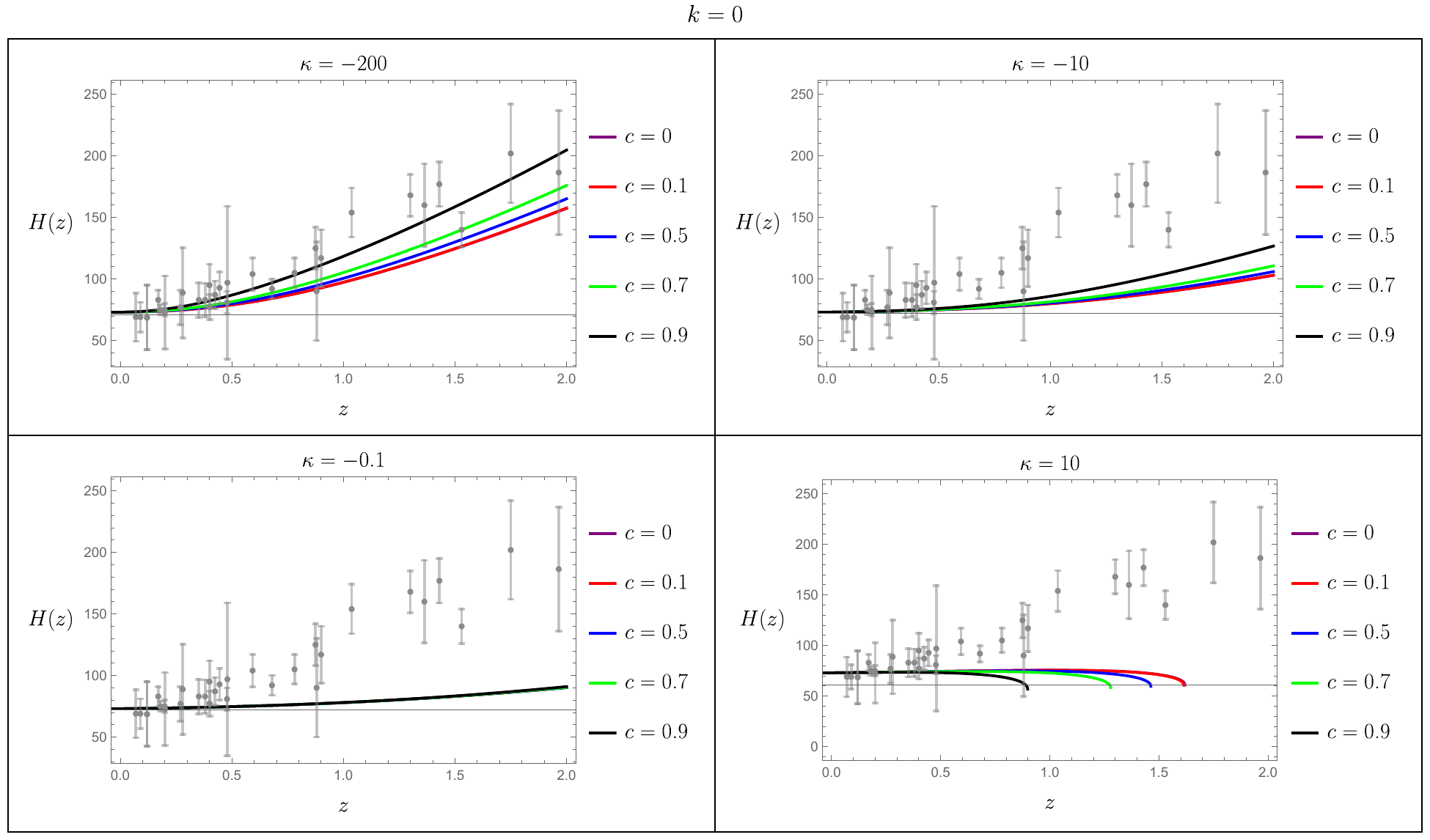}
	\caption{The figure shows numerical solutions of $w_\text{eff}(z) $ and $H(z)$ for $n=2$, that is $V = V_0 \phi^2$ 
		potential where $V_0 = 1/2$. The parameter $c$ is chosen to be $0, 0.1, 0.5, 0.7$ and $0.9$. NMDC coupling is $\k = -200, -10, -0.1, 10$. The spatial curvature is flat ($k=0$).  Considering $H(z)$ results obtained from \cite{Farooq:2016zwm}, the case of positive $\k$ is not favored while large negative value of $\k$, e.g. $\k = -200$ matches the observatioanl data better. Greater $c$, e.g. $c=0.9$ (black curve) results in greater speed of expansion. Considering $w_{\text{eff}}(z)$, greater negative $\k$ and greater $c$ allows phantom crossing at present time or at near future.}  \label{figwhflat} 
\end{figure}

The autonomous system can be integrated numerically. The flat case $w_{\text{eff}}(z)$ and $H(z)$ solutions are presented in figure \ref{figwhflat}. 
The numerical solutions are plotted confronting of the observed $H(z)$ error bar at low-$z$. The mean and error bar data used here is reported in \cite{Farooq:2016zwm}. These plots are to present the numerical results schematically in comparison to the data. They are without any statistical relevant between the numerical solutions of our model to the observational data.    To test and establish a statistical significance (for example, to solve the Hubble tension), more data from OHD+Pantheon+Masers should be necessary considered with the MCMC analysis. AIC and BIC analysis are to be performed for model selection.
As seen in the figure \ref{figwhflat}, positive NMDC coupling $\k$ neither gives any acceptable results for $w_{\text{eff}}(z)$ nor $H(z)$. Negative $\k$ is favored by the data. However, the negative NMDC coupling needs to be large, e.g. $\k = -200$ in the unit of $8 \pi G \equiv 1$, in order to schematically agree with the $H(z)$ data and its error bar.  Larger value of $c$ is proportional to larger holographic vacuum energy density. Therefore $c$ enhances both slope and magnitude of the $w_{\text{eff}}(z)$ and $H(z)$. Large negative NMDC coupling together with large fraction, $c$, of holographic vacuum density  can both enhance phantom effect. If the negative NMDC coupling is sufficiently strong, $w_{\text{eff}}(z)$ can be in phantom region as seen in figure \ref{figwhflat}.    

\subsection{Non-flat Case}
\begin{figure}[h]
	\centering
	\includegraphics[width=9.5cm]{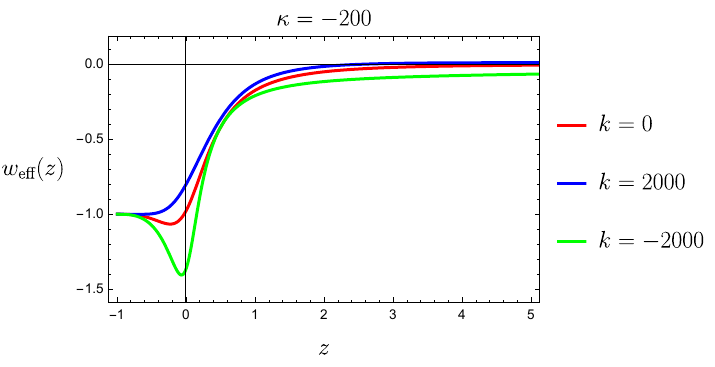}
	\includegraphics[width=9.5cm]{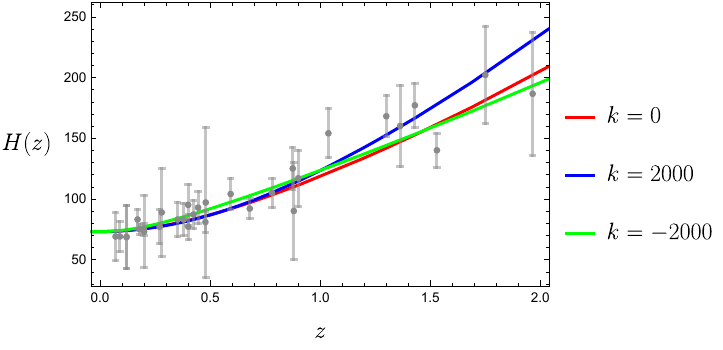}
	\caption{Numerical solution for the equation of state parameter $w_\text{eff}(z) $ and the Hubble rate $H(z)$ in case of flat and non-flat geometry for $n=2$, i.e. $V = V_0 \phi^2 $ potential where $V_0 =1/2$. The parameter $c$ is $0.9$ and the NMDC coupling is $\k = -200$. The spatial curvature is set to $k = 0, \pm 2000$.}  \label{figwhcompare} 
\end{figure}

Solutions for non-flat case are slightly different from those of the flat case. It is hard to tell difference between results of the flat case and the non-flat case.  In order to check qualitative effect of the spatial curvature, numerical value of $k$ is chosen to be large, i.e. $k = \pm 2000$, in order to magnify contribution of the curvature term that could affect $w_{\text{eff}}(z)$  and $H(z)$. This is shown in figure \ref{figwhcompare}. As seen in the figure, the  $w_{\text{eff}}(z)$  and $H(z)$ curves of the three cases ($k = 0, k > 0, k < 0$) are possible to cross each others. The crossing  is due to the coupling between $\k$ and $k$ in the Friedmann equations \eqref{fff1} and  \eqref{Ray}.  In the equation \eqref{fff1},  it is clear that NMDC terms with $\k > 0$ and $k > 0$ can contribute to phantom equation of state,  $w_{\text{eff}}(z) < -1$.  This is possible also with  $\k > 0$ and $k < 0$ (when $|k|/a^2 < 3 H^2$).
On the other hand, for the case $\k < 0$, this is possible only when $k < 0$ with $|k|/a^2 > 3 H^2$.
Since $	w_\text{eff} =   {[-1     - ({2\dot{H}}/{3H^2})    +    ({\Omega_k}/{3}) ]}/{(1-\Omega_k)}$, in order to have $ w_{\text{eff}}(z) < -1$, we need $\dot{H} > 0$.
For $k > 0$, the denominator is greater than $1$, reducing the phantom contribution. If  $k < 0$, the denominator is less than $1$, the phantom condition is  $2 \dot{H} > |k|/a^2$.

The holographic term with apparent horizon cutoff can contribute to   $w_{\text{eff}}(z) < -1$. Let us consider equation \eqref{rhoah}, $\rho_\Lambda =   \l[{3c^2}/{(8\pi G)}\r]\left(H^2+{k}/{a^2}\right) $. One can see that sign of $k$ could enhance or reduce rate of change the holographic vacuum density. Since in our consideration, $\rho_{\Lambda}$ is not constant and the continuity equation, $\dot{\rho}_{\Lambda} + 3 H {\rho}_{\Lambda}(1 + w_{\Lambda}) = 0$ (where $P_{\Lambda} = w_{\Lambda} \rho_{\Lambda}$) can read 
$
w_{\Lambda} = - 1  -  {\dot{\rho}_{\Lambda}}/({3 H {\rho}_{\Lambda}})
$
or
\be
\dot{\rho}_{\Lambda}   =   \f{6Hc^2}{8 \pi G} \l( \dot{H} - \f{k}{a^2} \r).
\ee
This gives 
\be
w_{\Lambda} = -1 -\f{2}{3 
	\l(H^2  +  k/a^2 \r)} \l(  \dot{H}  - \f{k}{a^2}  \r).
\ee
In order to have $w_{\Lambda} < -1$, if $k > 0$, the condition $\dot{H}  >  k/a^2$ is necessary. 
In case of $k < 0$, there are two subcases to consider, i.e. the case, $H^2 + k/a^2 > 0$ and the case  $H^2 + k/a^2 < 0$.  In order to have $w_{\Lambda} < -1$,  for $k < 0$ with $H^2 + k/a^2 > 0$, the condition $\dot{H}  >  k/a^2$ is necessary. For $k < 0$ with $H^2 + k/a^2 < 0$, the condition $\dot{H}  <  k/a^2$ is necessary, that $\dot{H}$ is negative. 
It is obvious from equation \eqref{fff1} and \eqref{Ray} that the NMDC terms (the term multiplied with $\k$) can contribute to phantom equation of state.  
Here, we can see that, not only the NMDC terms with $\k < 0$ that contribute to phantom equation of state, but free spatial curvature, $k$ terms (with large value of $k$) could contribute to phantom equation of state as well. The mixed NMDC and holographic effects to phantom equation of state is presented in figure \ref{figwhcompare} confronting with $H(z)$ data from \cite{Farooq:2016zwm}.

\section{Discussion and Conclusion}  \label{seccon} 
We consider an FLRW universe with arbitrary curvature. The cosmic matter contents are dust, NMDC scalar field driven by a power-law potential and holographic vacuum energy. The holographic IR cutoff length scale considered here is the apparent horizon which reduces to Hubble length when the space is flat. The apparent horizon is a plausible cutoff because in an accelerating universe, the horizon forms a trapped null surface in the same spirit with blackhole's event horizon. For the flat case, one can write an effective gravitational constant for Friedmann equation. However, as a generic case, the effective gravitational constant cannot be expressed for the non-flat cases. This is because there is a coupling between spatial curvature and NMDC coupling term. 
Moreover, effective gravitational constant cannot be expressed by factorization at the Lagrangian level. This is unlike the case of non-minimal coupling (NMC) theory.
 Therefore the holographic vacuum density is expressed using standard gravitational constant, not the effective one as it is in the flat case. 
We perform dynamical and stability analysis of this system and found that there are four independent fixed points. One fixed point (the point (b)) is a stable node corresponding to $w_{\text{eff}} = -1$ which exists for any value of the NMDC coupling, $\k$, however observation favors only the case $\k < 0$.     
The other fixed points are either unstable or saddle nodes. Cosmological implications of all fixed points are considered in this work. One branch of the stable fixed point (b) solution corresponds to  de-Sitter expansion with $ a(t) = {a_0} e^{\sqrt{\lambda_1}t}$.   
We perform numerical integration of the dynamical system and plot the result confronting $H(z)$ data from \cite{Farooq:2016zwm}. It is seen  that for flat universe, $H(z)$ observational data schematically agrees with large negative value of $\k$, i.e. $\k \approx -200$ whereas greater value of $c$ lifts up slope of the $H(z)$ plots. For positive $\k$ or small negative $\k$, the numerical results match the $H(z)$ data only at low redshifts. Larger magnitude of $c$ increases the slopes of both $w_{\text{eff}}$ and $H(z)$ curves. Larger value of the negative NMDC coupling, $\k$ and larger value of $c$ could contribute to phantom equation of state, $w_{\text{eff}} < -1$ in the near future, i.e. at small negative redshifts. With inclusion of the spatial curvature, there is NMDC-spatial curvature coupling term which can contribute to  sign of kinetic scalar energy according to the sign of $k$.  Negative $\k$ is favored since it contributes to larger value of scalar field kinetic term in the Friedmann equation.
That is to say, this NMDC-spatial curvature coupling could affect the phantom energy contribution. Moreover, free spatial curvature term, in holographic vacuum density cutoff and free spatial curvature term in the equation \eqref{fff1} and \eqref{Ray} can as well contribute to phantom equation of state. This is significant only when magnitude of $k$ is large. 
  We learn from this work that, in the non-flat case,  the gravitational constant cannot be considered as effective gravitational constant 
  and it does not appear at Lagrangian level. We also learn that phantom effect could be contributed with large negative spatial curvature, not only with the NMDC term or $c$. In non-holographic limit, $c=0$ and flat case, our model with observational data favors negative $\k$ for the late-time NMDC gravity in agreement with the early-universe inflationary constraints of the NMDC gravity  reported by Tsujikawa \cite{Tsujikawa:2012mk}. Dynamics of our model in non-holographic limit gives concordant results, i.e. fixed point solutions,  to the work reported by Sushkov and Galeev \cite{Sushkov:2023aya}. Regarding the current issue of Hubble tension which debates discrepancy of the Hubble parameters analyzed from CMB data 
  ($H_0 \approx 67\, \rm{km \cdot sec^{-1} \cdot Mpc^{-1}}$) and the late-local universe observations  ($H_0 \approx 73\, \rm{km\cdot sec^{-1}\cdot Mpc^{-1}}$)  \cite{Efstathiou:2013via, DiValentino:2021izs}, the initial condition given in our numerical integration is $H_0 = 73\, \rm{km\cdot sec^{-1}\cdot Mpc^{-1}}$ which is of the late-local universe.   Recent reports by \cite{Vagnozzi:2018jhn,Vagnozzi:2019ezj,Huang:2016fxc,Escamilla:2023oce,Vagnozzi:2023nrq,Gangopadhyay:2022bsh,Gangopadhyay:2023nli} suggest that combination of local expansion and the CMB data prefers phantom equation of state.  
  This could be possible if the initial condition of the numerical integration is lowered towards $H_0 \approx 67\, \rm{km\cdot sec^{-1}\cdot Mpc^{-1}}$ for the case of $\k = -200$ and high value of $c$, e.g. $c = 0.7$ to $0.9$ for both flat case ($k =0$) and open case ($k < 0$).  
  Our work cannot significantly address the solution to the Hubble tension. To test our model, more data from, for example, OHD+Pantheon+Masers is needed with MCMC analysis. Model selection is to be performed with AIC and BIC analysis.  At last, we notice that large negative $\kappa$ is super-Planckian. This could be effectively possible if one considers negative $\kappa=\kappa(\phi)$ to scale with $\phi^{-2}$, that is the NMDC term is re-scaled with $ \phi^{-2}$ as in \cite{Gumjudpai:2016frh} motivated from re-scaling invariant of the Horndeski Lagrangian  \cite{Bettoni:2013diz}. For a power-law potential, late-time small-field value enlarges the $\phi^{-2}$ factor such that effectively large NMDC coupling is attained.




\section*{Acknowledgements}
We thank Nandan Roy for useful discussion. This research project has been funded by Mahidol University (Fundamental Fund: fiscal year 2024 by National Science Research and Innovation Fund (NSRF)). AT  is supported by a RGJ-PhD scholarship. We are grateful to the referee for very useful suggestion on further observational data confrontation.


\end{document}